\documentclass[final]{aipproc}

\usepackage{graphicx,amsmath,amssymb,amsthm,color}

\def \d {{\rm d}}

\newtheorem*{theorem}{Theorem}

\layoutstyle{6x9}

\begin{document}

\title{Apparent horizons in D-dimensional Robinson--Trautman spacetime}
\author{Otakar Sv\'{\i}tek}{address={Institute of Theoretical Physics, Charles University in Prague, Faculty of Mathematics and Physics, V~Hole\v{s}ovi\v{c}k\'ach 2, 180~00 Praha 8, Czech Republic }}

\begin{abstract}
We derive the higher dimensional generalization of Penrose--Tod equation describing apparent horizons in Robinson--Trautman spacetimes. New results concerning the existence and uniqueness of its solutions in four dimensions are proven. Namely, previous results of Tod \cite{tod} are generalized to nonvanishing cosmological constant.
\end{abstract}

\maketitle

\section{Robinson--Trautman spacetime in D dimensions}

Robinson--Trautman spacetimes (containing aligned pure radiation or vacuum with a cosmological constant $\Lambda$) in any dimension were obtained by \cite{podolsky-ortaggio} using the geometric conditions of the original articles about the four-dimensional version of the spacetime \cite{RobinsonTrautman:1960,RobinsonTrautman:1962}. Namely, they required the existence of a twistfree, shearfree and expanding null geodesic congruence. They have arrived at the following metric valid in higher dimensions
\begin{equation}
 \d s^2=\frac{r^2}{P^2}\,\gamma_{ij}\,\d x^i\d x^j-2\,\d u\d r-2H\,\d u^2 
\end{equation}
where $2H=\frac{{\cal R}}{(D-2)(D-3)}-2\,r(\ln P)_{,u}-\frac{2\Lambda}{(D-2)(D-1)}\,r^2-\frac{\mu(u)}{r^{D-3}}$. The unimodular spatial $(D-2)$-dimensional metric $\gamma_{ij}(x)$ and the function $P(x,u)$ must satisfy the field equation ${\cal R}_{ij}=\frac{{\cal R}}{D-2}h_{ij}$ (with $h_{ij}=P^{-2}\gamma_{ij}$ being the rescaled metric). In $D=4$ the field equation is always satisfied and ${\cal R}$ (Ricci scalar of the metric $h$) generally depends on $x^i$. However, in $D>4$ the dependence on $x^i$ is ruled out (${\cal R}={\cal R}(u)$). But generally, it still allows a huge variety of possible spatial metrics $h_{ij}$ (e.g., for ${\cal R} > 0$ and $5 \leq D-2 \leq 9$ an infinite number of compact Einstein spaces were classified).

\section{Apparent horizon}
Event horizon is a global characteristic and therefore the full spacetime evolution is necessary in order to localize it. Therefore, over the past years different quasi-local characterization of black hole boundary were developed. The most important ones being apparent horizon \cite{hawking-ellis}, trapping horizon \cite{hayward} and isolated or dynamical horizon \cite{ashtekar}. The basic {\it local} condition in the above mentioned horizon definitions is the same: these horizons are sliced by marginally trapped surfaces with vanishing expansion of outgoing (ingoing) null congruence orthogonal to the surface.

In our case we will be dealing only with the condition of vanishing expansion. For the historical reasons and because it was already used in \cite{chow-lun} we will call the horizon apparent. Concretely, we will search for the past apparent horizon. Since in $D=4$ the solutions of the Robinson--Trautman equation are generally diverging when approaching $u=-\infty$ it is not possible to extend the spacetime to past null infinity. Therefore we cannot use the event horizon. In figure, the schematic conformal picture of Robinson--Trautman spacetime (for $D=4$ and without cosmological constant) is presented together with the approximate location of the horizons.

\begin{figure*}[h]
\includegraphics{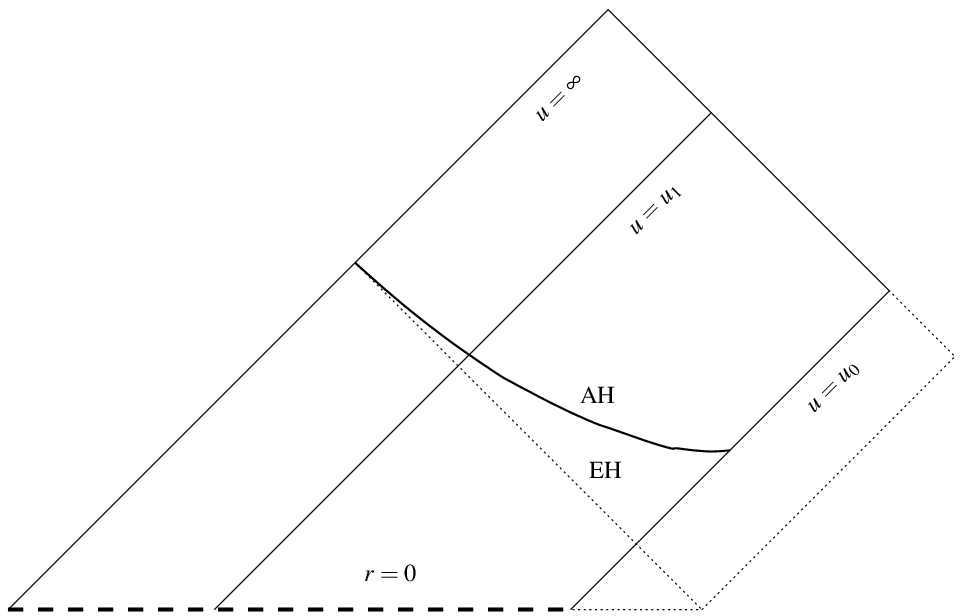} 
\end{figure*}

The explicit parametrization of the {\em past apparent horizon} hypersurface is $r=R(u,x^{i})$ such that its intersection with each $u=u_{1}$ slice is an outer marginally past trapped ($D-2$)-surface.

For the calculation of the expansion of an appropriate null congruence we will use a straight-forward generalization of the tetrad formalism to arbitrary dimension. Note that one can no longer use complex vector notation. Using two null covectors $l_{a}, n_{a}$ (with normalization $l_{a} n^{a}=-1$) and $D-2$ spatial covectors $m_{a\{i\}}$ ($i=1,..,D-2$) we suppose the following decomposition of the metric
\begin{equation}
g_{ab}=-2\,l_{(a}n_{b)}+m_{a\{i\}}m_{b\{j\}}\,\delta^{ij}
\end{equation}
Null D-ad adapted to the trapped hypersurface (using the above mentioned parametrization) has the following form:
\begin{equation}
l^{a}=(0,1,0,..,0);\ n^{a}=(1,\,[-H+{\textstyle \frac{1}{2}}g^{ij}R_{,i}R_{,j}]\, ,{\mathbf \nabla} R);\ m^{a}_{\{i\}}=(0,\,{\textstyle \frac{P}{r}}R_{,i}\,,{\textstyle \frac{P}{r}}{\mathbf w}_{i})
\end{equation}
where $D-2$ vectors ${\bf w}_{i}$ diagonalize metric $h$ and ${\bf \nabla} R=\{R^{,x^{1}},..,R^{,x^{D-2}}\}$. Fortunately, in subsequent calculations we do not need the explicit form of the vectors ${\bf w}_{i}$, it is sufficient to know their orthogonality properties.

By straight-forward computation one easily calculates the expansion associated with the congruence generated by $l^{a}$ to be $\Theta_{l}= \frac{D-2}{r}$ meaning that the outgoing null congruence is diverging. This is exactly what one assumes when dealing with the past trapped surface.

\section{Generalized Penrose--Tod equation}
Ingoing null congruence expansion can be calculated using the formula (sometimes a $(D-2)$ factor is used in the definition, but we are going to evaluate it to zero anyway) $\Theta_{n}=n_{a;b}\,p^{ab}$, where the tensor $p^{ab}=g^{ab}+2\,l^{(a}n^{b)}$ corresponds to the hypersurface projector. From $\Theta_{n}= 0$ (called Penrose--Tod equation in four dimensions) we get the trapped surface condition
\begin{equation}\label{GPT}
{\cal R}-{\frac{2(D-3)}{D-1}}\Lambda R^{2}-{(D-2)(D-3)}\frac{\mu}{R^{D-3}}-{2(D-3)}\Delta(\ln R)-
\end{equation}
\begin{equation*}
-{(D-4)(D-3)}(\nabla \ln R)\cdot (\nabla \ln R)= 0
\end{equation*}
It is a nonlinear PDE, where both the Laplacian and scalar product in the last term correspond to the Einstein metric $h_{ij}$. Interesting property of this equation is that for $D>4$ its nonlinearity is much worse since the term quadratic in derivatives appears.
\vskip 0.5cm
\noindent{\AIPsubsectionfont{\bf ${\mathbf{D=4:}}$ Existence of the solution}}\\
In four-dimensional case one can no longer use the existence proof given by Tod \cite{tod} when the cosmological constant is present. We will use the version of sub and super-solution method adapted to Riemannian manifolds given by Isenberg \cite{isenberg}. 
\begin{theorem}
{{\bf \emph{Sub and Super-solution method for equation} $\Delta \psi=f(x,\psi)$}}\newline
Let $\Sigma$ be a compact Riemannian manifold without boundary, and let $f: \Sigma\times {\mathbb R}_{+}\to {\mathbb R}$ be a smooth function. Assume that there exist functions $\phi_{-},\phi_{+}: \Sigma \to {\mathbb R}_{+}$ such that:
\begin{enumerate}
\item[$\bullet$] $0<\phi_{-}<\phi_{+},\ \Delta\phi_{-}\geq f(x,\phi_{-}),\ \Delta\phi_{+}\leq f(x,\phi_{+})$
\end{enumerate}
then there exists a function $\phi: \Sigma \to {\mathbb R}_{+}$ satisfying:
\begin{enumerate}
\item[$\bullet$] $\phi_{-}<\phi<\phi_{+},\ \Delta\phi=f(x,\phi)$
\end{enumerate}
\end{theorem}
\noindent Using the substitution $R=C\,e^{-\phi}$ ($C>0$) in equation (\ref{GPT}) we obtain
\begin{equation}
\Delta\phi=-\frac{{\cal R}}{2}+\frac{\Lambda}{3}C^{2}e^{-2\phi}+\frac{\mu}{C}e^{\phi}
\end{equation}
Now the equation for horizon has the form appropriate for the application of the theorem. To apply the theorem it is necessary to find the sub and super-solutions. The easiest choice is to look for the constants (making the left-hand side zero) that has to satisfy $0 \geq f(x,\phi_{-})$ and $0 \leq f(x,\phi_{+})$. We divide the cases according to the cosmological constant value:
\begin{enumerate}
\item ${{\Lambda \leq 0}:}$
Suppose $\phi_{min}>0$ (it can be always arranged by selecting high enough value of $C$), then $\phi_{-}=\ln\left(\frac{C}{2\mu}{\cal R}_{min}\right)$ and $\phi_{+}=\ln\left(\frac{C}{2\mu}{\cal R}_{max}-\frac{\Lambda}{3\mu}C^3\right)$ satisfy the conditions of the theorem if we choose $C>\frac{2\mu}{{\cal} R_{min}}$. This last condition is consistent with the previous demand that  $\phi_{min}>0$.
\item ${{\Lambda > 0}:}$ In this case we can satisfy the conditions only when $\Lambda < \frac{4}{9\mu^2}$ and  ${\cal R}_{min}<2$. For Schwarzschild--de-Sitter the first condition means an under-extreme case, which is correct restriction since the over-extreme one is naked.
\end{enumerate}
{\AIPsubsectionfont\bf ${\mathbf{D=4:}}$ Uniqueness} \\
For the proof of uniqueness we use the modification of Tod's proof incorporating the cosmological constant. Suppose $R_{1}$ and $R_{2}$ are solutions of (\ref{GPT}), subtract the corresponding equations for $R_{1}$ resp. $R_{2}$ (introducing $V=\frac{R_{1}}{R_{2}}$) to obtain
\begin{equation}\label{subtracted}
\Delta \ln V=-\frac{\mu}{R_{1}}(1-V)+\frac{\Lambda}{3}R_{2}^{2}(1-V^2)
\end{equation}
Multiplying equation (\ref{subtracted}) by $(1-V)$ and integrating it over the compact spatial surface (here we use the assumption that these surfaces are diffeomorphic to $S_{2}$) we get
\begin{equation}
-\int_{\Sigma}\left(\frac{\mu}{R_{1}}(1-V)^2-\frac{\Lambda}{3}R_{2}^{2}(1+V)(1-V)^2\right)=\int_{\Sigma}\frac{|\nabla V|^2}{V}
\end{equation}
Analysing the signs of both sides of this equation we have the following conclusions
\begin{enumerate}
\item For $\Lambda \leq 0$ the signs are opposite and so the only possibility is $V=1$ implying uniqueness.
\item For $\Lambda > 0$ we obtain opposite signs among the solutions satisfying $R \leq \sqrt[3]{\frac{3\mu}{2\Lambda}}$. It means that solution fulfilling this condition is only one. Interestingly, for extreme Schwarzschild--de-Sitter ($9\Lambda m^2=1$) we have $R \leq 3m$. One can then argue that this proves the uniqueness for the lower of both horizons.
\end{enumerate}

\section{Conclusion}
We have seen that existence and uniqueness results for the Penrose--Tod equation given by Tod can be generalized to nonvanishing cosmological constant. The limitations arising for positive $\Lambda$ are shown to be naturally related to the more complicated horizon structure of relevant spacetimes.

\begin{theacknowledgments}
This work was supported by grant GACR 202/07/P284 and the Center of Theoretical Astrophysics LC06014.
\end{theacknowledgments}

\bibliographystyle{aipproc}

\end{document}